# Indirect methods for wake potential integration

Igor Zagorodnov

*Deutsches Elektronen-Synchrotron (DESY), Notkestrasse 85, 22603 Hamburg, Germany*

May 30, 2006

The development of the modern accelerator and free-electron laser projects requires to consider wake fields of very short bunches in arbitrary three dimensional structures. To obtain the wake numerically by *direct* integration is difficult, since it takes a long time for the scattered fields to catch up to the bunch. On the other hand no general algorithm for *indirect* wake field integration is available in the literature so far. In this paper we review the known indirect methods to compute wake potentials in rotationally symmetric and cavity-like three dimensional structures. For arbitrary three dimensional geometries we introduce several new techniques and test them numerically.

# I. INTRODUCTION

The fast growth of computer power allows for direct time domain calculations of short-range wake potentials for general three dimensional accelerator elements. However, for short bunches a long-time propagation of the electromagnetic field in the outgoing vacuum chamber is required in order to take into account the scattered fields reaching the bunch later. To reduce drastically the computational time and to avoid the numerical error accumulation several indirect integration algorithms were developed for rotationally symmetric geometries [3-7]. For the general case in three dimensions such an algorithm is known only for cavity-like structures [8]. In this paper we review the known methods and introduce new techniques which allow for a treatment of arbitrary three dimensional structures. Several numerical examples are presented to illustrate the accuracy and efficiency of the described methods.

# II. FORMULATION OF THE PROBLEM

At high energies the particle beam is rigid. To obtain the electromagnetic wake field, the Maxwell equations can be solved with a rigid particle distribution [1, 2]. The influence of the wake field on the particle distribution is neglected here; thus, the beam-surrounding system is not solved self-consistently and a mixed Cauchy problem for the situation shown in Fig. 1 should be considered.

The problem reads as follows. For a bunch moving at velocity of light $c$ and characterized by a charge distribution $\rho$ find the electromagnetic field $\vec{E}, \vec{H}$ in a domain $\Omega$ which is bounded transversally by a perfect conductor $\partial \Omega$. The bunch introduces an electric current $\vec{j} = \vec{c}\rho$ and thus we have to solve for

$$\nabla \times \vec{H} = \frac{\partial}{\partial t}\vec{D} + \vec{j}, \qquad \nabla \times \vec{E} = -\frac{\partial}{\partial t}\vec{B}, \tag{1}$$

$$\nabla \cdot \vec{D} = \rho, \qquad \nabla \cdot \vec{B} = 0$$

$$\vec{H} = \mu^{-1}\vec{B}, \qquad \vec{D} = \varepsilon\vec{E}, \qquad x \in \Omega$$

$$\vec{E}(t=0) = \vec{E}_0, \qquad \vec{H}(t=0) = \vec{H}_0, \qquad x \in \overline{\Omega},$$

$$\vec{n} \times \vec{E} = 0, \qquad x \in \partial\Omega,$$

where $\vec{E}_0, \vec{H}_0$ is an initial electromagnetic field in the domain $\overline{\Omega}$ and $\vec{n}$ denotes the normal to the surface $\partial\Omega$.

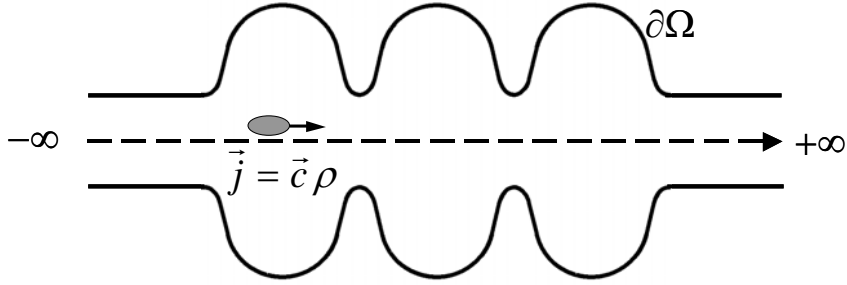

Fig. 1. Charged particle bunch moving through an accelerating structure supplied with infinite pipes.

The numerical methods to solve this problem were developed in [8-16].

We define the longitudinal and transverse wake potentials as [1, 2]

$$W_\parallel(\vec{r}, s) = -\frac{1}{Q} \int_{-\infty}^{\infty} E_z(\vec{r}, z, t(z,s)) dz, \qquad (2)$$

$$\vec{W}_\perp(\vec{r}, s) = \frac{1}{Q} \int_{-\infty}^{\infty} \left( \vec{E}_\perp + \vec{v} \times \vec{B} \right)(\vec{r}, z, t(z,s)) dz, \qquad (3)$$

where Q is the total charge of the bunch, $s$ is the distance behind the given origin $z_0 = ct$ in the exciting bunch, and

$$t(z,s) = (z+s)/c.$$

The purpose of this paper is to show how to replace the improper integrals in (2), (3) by proper integrals. This is essential for computer calculations, in particular for short bunches, where long beam tubes would require excessive computer memory and CPU time.

In the following only integral (2) will be considered. The transverse potential can be

derived from the longitudinal one by applying the Panofsky-Wenzel theorem [17]

$$\frac{\partial}{\partial s}\vec{W}_\perp(s,\vec{r}) = \nabla_\perp W_\parallel(s,\vec{r}). \tag{4}$$

### III. INDIRECT METHIODS FOR AXISYMMETRIC STRUCTURES

For rotationally symmetric structures, an azimuthal Fourier expansion can be used to reduce the problem to a set of two dimensional problems. For cavity like structures the integration of the wake fields can be performed along a straight line parallel to the axis at the outgoing beam tube radius as was suggested by T.Weiland in [4] and realized in codes BCI [10], TBCI [11] and MAFIA [13]. However, this technique works only if no part of the structure extends to a radius smaller than the radius of the outgoing tube. It has been realized later [5, 6] that the potential can be calculated by integrating the wake along the perfectly conducting boundary of a structure. Finally, O. Napoly et al [7] have generalized the above results by showing that the wake potentials, at all azimuthal harmonics $m$, can be expressed as integrals over the wake fields along any arbitrary contour spanning the structure longitudinally. This general method was implemented and tested in code ABCI [12]. A modified version of this method was introduced in paper [14] and implemented in code Echo.

An alternative approach based on waveguide mode expansion was introduced in [3] and realized in code DBCI.

In the following we review the most simple and general method of Napoly et al. and describe its modified version used later for the 3D case.

### A. Napoly-Chin-Zotter (NCZ) method for arbitrary rotational symmetric structures with unequal beam tubes radii

In this paper we consider only structures supplied with perfectly conducting ingoing and outgoing waveguides. The steady-state field pattern of a bunch in an ingoing perfectly conducting waveguide does not contribute to the wake potential. Hence the improper integral for the ingoing waveguide reduces to a proper integral along a finite part of the integration path and, as will be described below, the NCZ method is applicable for the case where the ingoing and outgoing tubes have unequal radii (see Fig. 2).

For a bunch moving at speed of light $c$ at an offset $r_0$ from and parallel to the axis of a rotationally symmetric structure, the source current $\vec{j}$ can be presented as

$$\vec{j} = \frac{\vec{c}Q\lambda(z/c-t)\delta(r-r_0)}{\pi a}\sum_{m=0}^{\infty}\frac{\cos m\varphi}{1+\delta_{m0}}, \qquad \int_{-\infty}^{\infty}\lambda(s)ds = 1, \tag{5}$$

where $\lambda(s)$ is the normalized longitudinal charge distribution and $Q$ is the bunch charge.

The scattered electromagnetic field

$$\vec{E}^{sc} = \vec{E} - \vec{E}^0, \quad \vec{B}^{sc} = \vec{B} - \vec{B}^0$$

can be written as

$$\left(E_r^{sc}, B_\theta^{sc}, E_z^{sc}\right)(r,\theta,z,t(z,s)) = \sum_{m=0}^{\infty}\left(e_r, b_\theta, e_z\right)^{(m)}(r,z,s)\cos(m\theta) \tag{6}$$

$$\left(B_r^{sc}, E_\theta^{sc}, B_z^{sc}\right)(r,\theta,z,t(z,s)) = \sum_{m=0}^{\infty}\left(b_r, e_\theta, b_z\right)^{(m)}(r,z,s)\sin(m\theta).$$

Substitution of expansion (6) in equations (1) and combining them yields [7] at each order $m$

$$\partial_r\left(r^m[e_z+cb_z]^{(m)}\right) = r^m\partial_z\left[e_r+cb_\theta-e_\theta+cb_r\right]^{(m)}, \tag{7}$$

$$\partial_r\left(r^{-m}[e_z-cb_z]^{(m)}\right) = r^{-m}\partial_z\left[e_r+cb_\theta+e_\theta-cb_r\right]^{(m)}. \tag{8}$$

That means that the 1-forms

$$\omega_S^{(m)} = r^m[e_r+cb_\theta-e_\theta+cb_r]^{(m)}dr + r^m[e_z+cb_z]^{(m)}dz, \tag{9}$$

$$\omega_D^{(m)} = r^{-m}[e_r+cb_\theta+e_\theta-cb_r]^{(m)}dr + r^{-m}[e_z-cb_z]^{(m)}dz$$

are closed.

Let us remind that the condition for 1-form $\omega = f(r,z)dr + g(r,z)dz$ to be closed is

$$\partial_r g - \partial_z f = 0 \quad \text{or} \quad d\omega = \left(\partial_r g - \partial_z f\right) dx \wedge dy = 0.$$

Hence, it follows from Stokes' theorem

$$\int_{\partial\Omega} \omega = \int_{\Omega} d\omega,$$

that an integral of the closed form along a closed contour vanishes. This property allows to deform the wake field integration path as described below.

For a perfectly conducting round pipe one can easily obtain [1]

$$e_r^{(m)}(z=\pm\infty) = cb_\theta^{(m)}(z=\pm\infty) = \begin{cases} \lambda(s)\dfrac{Q}{4\pi\varepsilon_0}\dfrac{1}{r}, & m=0 \\ \lambda(s)\dfrac{Q}{2\pi\varepsilon_0}\dfrac{r_0^m}{a^{2m}}r^{m-1}, & m>0 \end{cases},$$

$$-e_\theta^{(m)}(z=\pm\infty) = cb_r^{(m)}(z=\pm\infty) = \begin{cases} 0, & m=0 \\ \lambda(s)\dfrac{Q}{2\pi\varepsilon_0}\dfrac{r_0^m}{a^{2m}}r^{m-1}, & m>0 \end{cases}$$

$$\omega_D(z=\pm\infty) = 0,$$

where $a$ denotes the pipe radius.

The longitudinal wake potential at mode $m$ is defined as

$$W_\parallel^{(m)}(r,s) = -\frac{1}{Q}\int_{-\infty}^{\infty} e_z^{(m)}(r,z,s)dz. \tag{10}$$

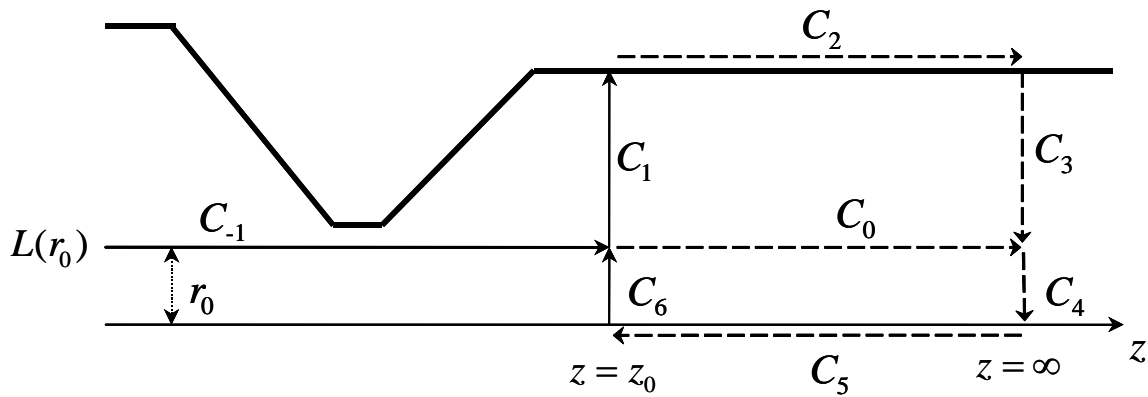

Fig. 2. Contours for the indirect integration.

Hence, for the general situation shown in Fig. 2 we can write [7]

$$QW_\|^{(m)} = -\int_{-\infty}^{\infty} e_z^{(m)} dz = -\int_{L(r_0)} e_z^{(m)} dz = -0.5 r_0^m \int_{L(r_0)} \left(\omega_D^{(m)} + a^{-2m}\omega_S^{(m)}\right), \quad L_{r_0} = C_{-1} \cup C_0, \qquad (11)$$

where we have used the relation

$$\int_{L(r_0)} \omega_S^{(m)} = 0.$$

This gives (we simplify the notation and omit the azimuthal number)

$$QW_\|^{(m)} = -0.5 r_0^m \int_{L(r_0)} \left(\omega_D + a^{-2m}\omega_S\right) = -0.5 r_0^m \int_{C_{-1}} \left(\omega_D + a^{-2m}\omega_S\right) + 0.5 r_0^m \int_{C_{13}} \left(\omega_D + a^{-2m}\omega_S\right),$$

$$C_{13} = \bigcup_{i=1}^{3} C_i.$$

For a perfectly conducting outgoing pipe we can write

$$\frac{r_0^m}{2} \int_{C_{13}} \left(\omega_D + a^{-2m}\omega_S\right) = \frac{r_0^m}{2} \int_{C_1} \left(\omega_D + a^{-2m}\omega_S\right) + \frac{r_0^m}{2a^{2m}} \int_{C_3} \omega_S =$$

$$= \frac{r_0^m}{2a^m} \int_{r_0}^{a} \left\{\left(\frac{a^m}{r^m} + \frac{r^m}{a^m}\right)[e_r + cb_\theta] + \left(\frac{a^m}{r^m} - \frac{r^m}{a^m}\right)[e_\theta - cb_r]\right\} + QF^{(m)}(s)$$

and the wake potential can be found as

$$W_\|^{(m)} = -\frac{r_0^m}{2Qa^m} \int_{C_{-1}} \left(\left[\frac{r_0^m}{a^m} + \frac{a^m}{r_0^m}\right] e_z - \left[\frac{r_0^m}{a^m} - \frac{a^m}{r_0^m}\right] cb_z\right) dz + \qquad (12)$$

$$+ \frac{r_0^m}{2Qa^m} \int_{r_0}^{a} \left\{\left(\frac{a^m}{r^m} + \frac{r^m}{a^m}\right)[e_r + cb_\theta] + \left(\frac{a^m}{r^m} - \frac{r^m}{a^m}\right)[e_\theta - cb_r]\right\} + F^{(m)}(s),$$

where

$$F^{(m)}(s) = \begin{cases} \dfrac{\lambda(s)}{2\pi\varepsilon_0} \ln\left(\dfrac{a}{r_0}\right), & m = 0, \\ \dfrac{\lambda(s)}{2\pi\varepsilon_0} \dfrac{r_0^{2m}}{ma^{4m}} (r_0^{2m} - a^{2m}), & m > 0. \end{cases}$$

Following the NCZ method we managed to replace the improper integration along the contour $C_0$ by the proper integral along the finite contour $C_1$.

## B. Modification of the NCZ method

In this paragraph we introduce a modification of the NCZ method. The main feature of our method is that (like in the direct method) we integrate only the $e_z^{(m)}(r,z,s)$ component of the scattered electromagnetic field along a straight line $C_{-1}$ at radius $r_0$, and use other field components only at the end of the structure. This property of the method allows to apply it for 3D calculations as described in section IV.B.

For the general situation shown in Fig. 2 we can write

$$QW_\parallel^{(m)} = -\int_{-\infty}^{\infty} e_z^{(m)} dz = -\int_{L(r_0)} e_z^{(m)} dz = -\int_{C_{-1}} e_z^{(m)} dz - \int_{C_0} e_z^{(m)} dz, \quad L_{r_0} = C_{-1} \cup C_0, \quad (13)$$

$$-\int_{C_0} e_z^{(m)} dz = -0.5 \left( \int_{C_0} \left( r_0^m \omega_D^{(m)} + r_0^{-m} \omega_S^{(m)} \right) + \frac{\beta}{a^m} \int_{C_{16}} \omega_S^{(m)} \right) =$$

$$= 0.5 \left( \int_{C_{13}} \left( r_0^m \omega_D^{(m)} + r_0^{-m} \omega_S^{(m)} \right) - \frac{\beta}{a^m} \int_{C_{16}} \omega_S^{(m)} \right),$$

$$\beta = \left(\frac{a}{r^0}\right)^m - \left(\frac{a}{r^0}\right)^{-m}, \quad C_{16} = \bigcup_{i=1}^{6} C_i .$$

For a perfectly conducting geometry the last equation reduces to

$$\int_{C_0} e_z^{(m)} dz = -0.5 \left( \int_{C_1} \left( r_0^m \omega_D^{(m)} + r_0^{-m} \omega_S^{(m)} - \frac{\beta}{a^m} \omega_S^{(m)} \right) - \frac{\beta}{a^m} \int_{C_6} \omega_S^{(m)} \right)$$

and the wake potential can be found as

$$QW_\parallel^{(m)} = -\int_{C_{-1}} e_z^{(m)} dz + \frac{\beta}{2a^m} \int_0^a r^m [e_r + cb_\theta - e_\theta + cb_r]^{(m)} dr - \quad (14)$$

$$-0.5 \int_{r_0}^a \left\{ \left(\frac{r_0^m}{r^m} + \frac{r_0^{-m}}{r^{-m}}\right) [e_r + cb_\theta]^{(m)} + \left(\frac{r_0^m}{r^m} - \frac{r_0^{-m}}{r^{-m}}\right) [e_\theta - cb_r]^{(m)} \right\} dr$$

Again we managed to replace the improper integration along the contour $C_0$ by proper integrals along the finite contours $C_1, C_6$.

## IV. INDIRECT METHODS FOR 3D STRUCTURES

In the previous chapter a general solution for rotationally symmetric geometries was described. However the NCZ method does not generalize to three dimensions. Following the same route it is easy to obtain closed 2-forms. However, wake potential (2) is a line integral and it cannot be treated through 2-forms. Hence, we have to look here for alternative methods.

### A. Method for cavity like structures or structures with small outgoing waveguide

As for the rotationally symmetric case the integration through a waveguide gap results in a simple and efficient algorithm [8].

As shown in [8] the longitudinal wake potential is a harmonic function of the transverse coordinates

$$\Delta_\perp W_\parallel(s, x, y) = 0, \quad (x, y) \in \Omega_{max}^\perp, \tag{15}$$

where $\Omega_{max}^\perp$ is the transverse area constituted by intersection of all transverse cross-sections (see, for example, Fig.4). Hence for cavity like structures the relation $\Omega_{out}^\perp = \Omega_{max}^\perp$ holds and we perform the integration at the transverse position of the outgoing waveguide boundary $\partial \Omega_{out}^\perp$. The longitudinal wake potential for any position inside the waveguide is then obtained as a solution of the Laplace equation (15) with Dirichlet boundary condition

$$W_\parallel(s, x, y) = W_\parallel^{direct}(s, x, y), \quad (x, y) \in \partial \Omega_{out}^\perp.$$

However, this method does not work if the area $\Omega_{max}^\perp$ is smaller than the outgoing waveguide intersection $\Omega_{out}^\perp$. Below we suggest methods able to treat this situation.

### B. Method for general 3D structures with outgoing round pipe

In this paragraph we consider the situation where an arbitrary three dimensional structure is supplied with a round outgoing pipe. In this case we can easily generalize our method (14) as follows.

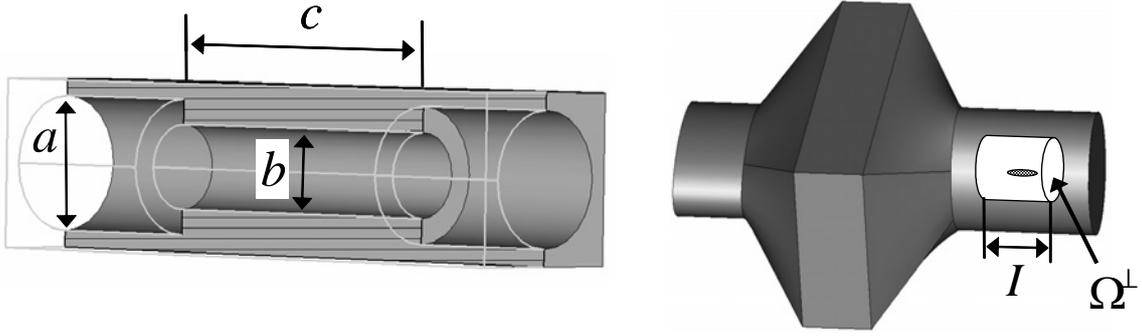

Fig. 3. The round step collimator and the tapered cavity. Area $\Omega^\perp \times I$ around the bunch for wakepotential calculation.

Let us suggest that we are interested in the wakepotential in area $\Omega^\perp \times I$ around the bunch as shown in Fig.3

$$W_\|(\vec{r},s) = -\frac{1}{Q}\int_{-\infty}^{\infty} E_z(\vec{r},z,t(z,s))dz, \quad \vec{r}=(x,y)\in\Omega^\perp,\ s\in I. \quad (16)$$

Let us the same as in rotationally symmetric case present the integration path $L(\vec{r}_0)$ as union of the path $C_{-1}(\vec{r}_0)$ through the 3D structure up to the point $z_0$ and the path $C_0(\vec{r}_0)$ completely inside of the outgoing round pipe

$$L(\vec{r}_0) = C_{-1}(\vec{r}_0) \cup C_0(\vec{r}_0), \qquad \vec{r}_0 \in \Omega^\perp.$$

The wake potential can be written as

$$QW_\|(\vec{r}_0,s) = -\int_{C_{-1}(\vec{r}_0)} E_z^{sc}(\vec{r}_0,z,t(z,s))dz - \int_{C_0(\vec{r}_0)} E_z^{sc}(\vec{r}_0,z,t(z,s))dz, \quad \vec{r}_0 \in \Omega^\perp \quad (17)$$

Our purpose is to replace the second improper integral by proper integrals. This can be achieved by straightforward generalization of the method described in section III.B.

Indeed, after the bunch arrived in the round pipe we can use an azimuthal Fourier expansion to reduce the 3D problem to set of 2D problems. However, unlike in the rotationally symmetric case of Section III the electromagnetic field components are now complex quantities due to the fields scattered before by the 3D structure.

Let us represent the scattered electromagnetic field $\vec{F} = (\vec{E}^{sc}, \vec{B}^{sc})$ as

$$\vec{F}(z,r,\theta) = \mathrm{Re} \sum_{m=0}^{\infty} \vec{F}_m(z,r) e^{-im\theta} = \mathrm{Re} \sum_{m=0}^{\infty} \left[ \vec{F}_m^{\mathrm{Re}}(z,r) + i\vec{F}_m^{\mathrm{Im}}(z,r) \right] e^{-im\theta},$$

$$\vec{F}_m(z,r) = 2\int_0^{2\pi} \vec{F}(z,r,\theta) e^{im\theta} \frac{d\theta}{2\pi}, \qquad \vec{F}_0(z,r) = \int_0^{2\pi} \vec{F}(z,r,\theta) \frac{d\theta}{2\pi}.$$

Then the equations for the azimuthal harmonics at each number $m$ separate into two independent sets and each of them can be written in the form (we simplify the notation and omit the azimuthal number)

$$\frac{1}{r}\frac{\partial}{\partial r}(rb_r) + \frac{M}{r}b_\theta + \frac{\partial}{\partial z}b_z = 0, \qquad \frac{1}{r}\frac{\partial}{\partial r}(re_r) + \frac{M}{r}e_\theta + \frac{\partial}{\partial z}e_z = 0,$$

$$\frac{M}{r}b_z - \frac{\partial}{\partial z}b_\theta = \frac{\partial}{c^2\partial t}e_r, \qquad -\frac{M}{r}e_z - \frac{\partial}{\partial z}e_\theta = -\frac{\partial}{\partial t}b_r,$$

$$\frac{\partial}{\partial z}b_r - \frac{\partial}{\partial r}b_z = \frac{\partial}{c^2\partial t}e_\theta, \qquad \frac{\partial}{\partial z}e_r - \frac{\partial}{\partial r}e_z = -\frac{\partial}{\partial t}b_\theta,$$

$$\frac{1}{r}\frac{\partial}{\partial r}(rb_\theta) - \frac{M}{r}b_r = \frac{\partial}{c^2\partial t}e_z, \qquad \frac{1}{r}\frac{\partial}{\partial r}(re_\theta) + \frac{M}{r}e_r = -\frac{\partial}{\partial t}b_z,$$

where $M = m$ for $e_z^{\mathrm{Re}}, e_r^{\mathrm{Re}}, b_\theta^{\mathrm{Re}}, b_z^{\mathrm{Im}}, b_r^{\mathrm{Im}}, e_\theta^{\mathrm{Im}}$ and $M = -m$ for $e_z^{\mathrm{Im}}, e_r^{\mathrm{Im}}, b_\theta^{\mathrm{Im}}, b_z^{\mathrm{Re}}, b_r^{\mathrm{Re}}, e_\theta^{\mathrm{Re}}$.

The same as in Section III.B we can show that the differential forms

$$\omega_S^{(M)} = r^M [e_r + cb_\theta - e_\theta + cb_r]^{(M)} dr + r^M [e_z + cb_z]^{(M)} dz,$$

$$\omega_D^{(M)} = r^{-M} [e_r + cb_\theta + e_\theta - cb_r]^{(M)} dr + r^{-M} [e_z - cb_z]^{(M)} dz,$$

are closed and the following relations hold

$$\int_{C_0} e_z^{\mathrm{Re},(m)} dz = -\frac{1}{2}\left( \int_{C_1}\left( r_0^m \omega_D^{(m)} + r_0^{-m} \omega_S^{(m)} - \frac{\beta}{a^m} \omega_S^{(m)} \right) - \frac{\beta}{a^m} \int_{C_6} \omega_S^{(m)} \right), \qquad (18)$$

$$\int_{C_0} e_z^{\mathrm{Im},(m)} dz = -\frac{1}{2}\left( \int_{C_1}\left( r_0^m \omega_S^{(-m)} + r_0^{-m} \omega_D^{(-m)} - \frac{\beta}{a^m} \omega_D^{(-m)} \right) - \frac{\beta}{a^m} \int_{C_6} \omega_D^{(-m)} \right).$$

The second integral in relation (17) can be written as

$$\int_{C_0(\vec{r}_0)} e(\vec{r}_0, z, s) dz = \sum_{m=0}^{\infty}\left( \int_{C_0} e_z^{\mathrm{Re},(m)} dz \cos(m\theta_0) + \int_{C_0} e_z^{\mathrm{Im},(m)} dz \sin(m\theta_0) \right), \quad \vec{r}_0 = (r_0, \theta_0) \in \Omega^\perp, (19)$$

and substitution of equations (18) reduces this improper integral along the z-axis to a sum of proper integrals along the radius.

### C. Method based on the directional symmetry of wake potential

The methods introduced in the previous sections are not fully general. The method of section IV.A allows to treat only structures where the crosssection of the outgoing waveguide is covered by any other crosssection along the structure. For example, if we are interested in the wake for the transition from a round pipe to a rectangular one, as shown in Fig.4, then this method does not work. The method of the section IV.B is not applicable directly, too.

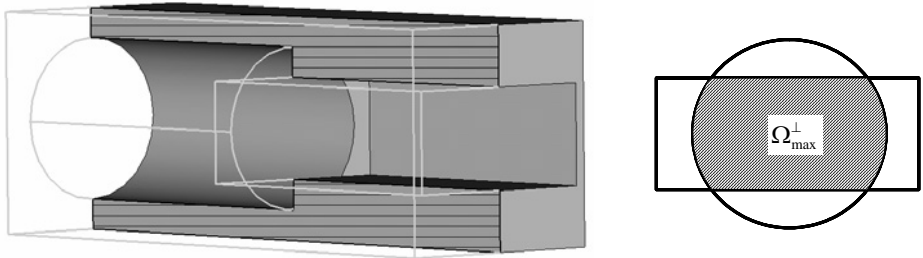

Fig. 4. Round to rectangular transition and the maximal area for wakepotential calculation.

However, often we are able to apply one of the two methods when the bunch direction of motion is reversed. For example, the inverse transition from a rectangular to a round pipe can be treated with the method of section IV.B.

In this section we describe a method which allows to calculate the wakepotential for one direction from the wakepotential for the reversed one.

In [18] a directional symmetry of the impedance

$$Z_\|(\omega,\vec{r}) = \int_{-\infty}^{\infty} W_\|(\tau,\vec{r}) e^{-i\omega\tau} d\tau$$

was considered and the relation between the forward impedance $Z^+(\omega,\vec{r}_e)$ and the "reversed" impedance $Z^-(\omega,\vec{r}_e)$ was found

$$Z^-(\omega, \vec{r}_e) - Z^+(\omega, \vec{r}_e) = \frac{2}{Q^2} \text{Re} \left( \int_{\Omega_{in}^\perp} \vec{E}^+ \times \vec{H}^- d\vec{\mu}(\vec{r}) - \int_{\Omega_{out}^\perp} \vec{E}^+ \times \vec{H}^- d\vec{\mu}(\vec{r}) \right),$$

where $\Omega_{in}^\perp$ denotes the ingoing and $\Omega_{out}^\perp$ is the outgoing pipe cross-section. However, in order to apply the Panowsky-Wenzel theorem and find the transverse wake potential we need to know the longitudinal wakepotential not only at the position of the bunch $\vec{r}_e$ but in some vicinity of it.

Below we generalize method of paper [18] in order to be able to calculate the transverse wakepotential, too.

Let us consider a perfectly conducting structure traversed by two point charges traveling parallel to the z-axis in opposite directions and with offsets $\vec{r}_1$ and $\vec{r}_2$, correspondingly. The current densities in frequency domain are

$$J_z^+(\vec{r}, z) = Q\delta(\vec{r} - \vec{r}_1) e^{-ikz}, \quad k = \omega/c,$$

$$J_z^-(\vec{r}, z) = -Q\delta(\vec{r} - \vec{r}_2) e^{ikz}.$$

From the Lorentz reciprocity theorem [20] we obtain

$$Z^-(\omega, \vec{r}_2, \vec{r}_1) - Z^+(\omega, \vec{r}_1, \vec{r}_2) = \frac{1}{Q^2} \left( \int_{\Omega_{in}^\perp} \vec{F}(\vec{r}_1, \vec{r}_2, \vec{r}) d\vec{\mu}(\vec{r}) - \int_{\Omega_{out}^\perp} \vec{F}(\vec{r}_1, \vec{r}_2, \vec{r}) d\vec{\mu}(\vec{r}) \right), \quad (20)$$

$$\vec{F}(\vec{r}_1, \vec{r}_2, \vec{r}) = \vec{E}^+(\vec{r}_1, r) \times \vec{H}^-(\vec{r}_2, r) - \vec{E}^-(\vec{r}_2, r) \times \vec{H}^+(\vec{r}_1, r).$$

In order to calculate the right-hand side we note that at infinity the field patern of the charge can be found by solving the two dimensional Poisson equation

$$\Delta \varphi_i(\vec{r}) = Z_0 Q \delta(\vec{r} - \vec{r}_i), \quad \vec{r} \in \{\Omega_{in}^\perp, \Omega_{out}^\perp\}, \quad (21)$$

$$\varphi_i(\vec{r}) = 0, \ \vec{r} \in \{\partial\Omega_{in}^\perp, \partial\Omega_{out}^\perp\}, \quad i = 1, 2.$$

To show this observe that electric fields at infinity can be written as

$$\vec{E}_1(x, y, z) = e^{-ikz} \nabla \varphi_1(x, y), \quad \vec{E}_2(x, y, z) = e^{ikz} \nabla \varphi_2(x, y). \quad (22)$$

Substituting this representation into Maxwell's equation

$$\nabla \times \nabla \times \vec{E}_i = k^2 \vec{E}_i - i\omega\mu \vec{J}_i$$

yields equation (21). Additionally, from representations (22) and Maxwell's equation

$$\nabla \times \vec{E}_i = -i\omega\mu \vec{H}_i,$$

we obtain that at infinity the following relations hold

$$\vec{H}_1 = \frac{1}{Z_0} \vec{e}_z \times \vec{E}_1, \qquad \vec{H}_2 = -\frac{1}{Z_0} \vec{e}_z \times \vec{E}_2,$$

where $Z_0$ is free space impedance.

Hence, equation (20) can be written as

$$Z^-(\omega, \vec{r}_2, \vec{r}_1) - Z^+(\omega, \vec{r}_1, \vec{r}_2) = 2Z^e(\vec{r}_1, \vec{r}_2), \tag{23}$$

$$Z^e(\vec{r}_1, \vec{r}_2) = \frac{1}{Q^2 Z_0} \left( \int_{\Omega_{in}^\perp} (\nabla \varphi_1, \nabla \varphi_2) d\mu(\vec{r}) - \int_{\Omega_{out}^\perp} (\nabla \varphi_1, \nabla \varphi_2) d\mu(\vec{r}) \right)$$

and forward and reverse wakepotentials are related as

$$W^-(s, \vec{r}_2, \vec{r}_1) - W^+(s, \vec{r}_1, \vec{r}_2) = 2W^e(s, \vec{r}_1, \vec{r}_2), \tag{24}$$

$$W^e(s, \vec{r}_1, \vec{r}_2) = \int_{-\infty}^{s} w_\delta^e(s-s') \lambda(s') ds' = c Z^e(\vec{r}_1, \vec{r}_2) \lambda(s),$$

$$w_\delta^e(s) = \int_{-\infty}^{\infty} Z^e(\vec{r}_1, \vec{r}_2) e^{i\frac{\omega}{c}s} \frac{d\omega}{2\pi} = c\delta(s) Z^e(\vec{r}_1, \vec{r}_2).$$

Hence, in order to find the wake potential of the round to rectangular transition we can calculate the wakepotential of the inverse transition with the method of section IV.B. Then we need to calculate fileds (22) in both pipe cross-sections. This is a two dimensional problem and can be solved either analytically or numerically. Finally, we use equation (24) to obtain the required wakepotential. The numerical application of this method for the calculation of the wakes can be found in [19].

**D. General method based on waveguide mode expansion**

In this section we present a general method for arbitrary 3D geometries. The method is based on a waveguide mode expansion. The first attempt to use a waveguide mode expansion was made in the code DBCI for monopole and dipole azimuthal modes in rotationally symmetric structures [3].

In this section we derive a method for the general three dimensional case. Compared to the considerations of [3] the resulting algorithm exhibits much simpler equations which are easily realized numerically.

The longitudinal component $E_z$ of the scattered electric field in the outgoing waveguide can be written as a linear combination of the z-components of the TM waveguide modes [20]

$$E_z^{sc}(\vec{r}_\perp, z, t) = \sum_n E_n(\vec{r}_\perp) \int_{-\infty}^{\infty} \alpha_n(\omega) e^{i(\beta_n z - \omega t)} d\omega = \sum_n E_n(\vec{r}_\perp) g_n(z,t).$$

In the general case we can again represent the wakepotential as

$$QW_\parallel(\vec{r}_0, s) = - \int_{C_{-1}(\vec{r}_0, z_0 - s)} E_z^{sc}(\vec{r}_0, z, t(z,s)) dz - \int_{C_0(\vec{r}_0, z_0 - s)} E_z^{sc}(\vec{r}_0, z, t(z,s)) dz, \quad t(z,s) = \frac{z+s}{c}.$$

The second integral can be written as

$$\int_{C_0(\vec{r}_0, z_0 - s)} E_z^{sc}(\vec{r}_0, z, t(z,s)) dz = \int_{z_0 - s}^{\infty} E_z^{sc}(\vec{r}_0, z, t = \frac{z+s}{c}) dz =$$

$$= \int_{z_0 - s}^{\infty} \sum_n E_n(\vec{r}_\perp) \int_{-\infty}^{\infty} \alpha_n(\omega) e^{i\left(\beta_n z - \omega \frac{z+s}{c}\right)} d\omega dz = -\sum_n E_n(\vec{r}_\perp) \int_{-\infty}^{\infty} \alpha_n(\omega) \frac{e^{i\left(\beta_n (z_0 - s) - \omega \frac{z_0}{c}\right)}}{i\left(\beta_n - \frac{\omega}{c}\right)} d\omega =$$

$$= -\sum_n \frac{E_n(\vec{r}_\perp)}{k_n^2} \int_{-\infty}^{\infty} \alpha_n(\omega) e^{i\left(\beta_n (z_0 - s) - \omega \frac{z_0}{c}\right)} i\left(\beta_n + \frac{\omega}{c}\right) d\omega,$$

where $k_n^2 = (\omega c^{-1})^2 - \beta_n^2$ is a squared cutoff wave number for mode $n$ (see Eq. (27)).

From the last expression we obtain

$$\int_{z_0 - s}^{\infty} E_z^{sc}(\vec{r}_0, z, t(z,s)) dz = \sum_n \frac{E_n(\vec{r}_\perp)}{k_n^2} \left[\frac{\partial}{\partial s} + \frac{\partial}{c \partial t}\right] g_n(z_0 - s, t_0), \qquad (25)$$

where $t_0 = z_0 c^{-1}$ and

$$g_n(z_0 - s, t_0) = \int_{\Omega_{out}^\perp} E_z^{sc}(x, y, z_0 - s, t_0) E_n(x, y) dx dy . \qquad (26)$$

Equation (25) represents the main result of this paper. It reduces the improper integral along the z-axis to the sum of proper integrals (26) in the transverse waveguide cross-section $\Omega_{out}^\perp$.

Let us describe shortly a numerical algorithm. As a first step we should find the eigenmodes $\{E_n(x, y)\}$ and the eigenvalues $\{k_n^2\}$ from the solution of the eigenvalue problem [20]

$$\Delta E_n(x, y) = -k_n^2 E_n(x, y), \qquad (x, y) \in \Omega_{out}^\perp . \qquad (27)$$

Next we have to find coefficient functions (26) at two instants of time: $t_0 - 0.5\Delta t$ and $t_0 + 0.5\Delta t$. Finally, we approximate equation (25) as

$$\int_{z_0 - s}^{\infty} E_z^{sc}(\vec{r}_0, z, t = \frac{z+s}{c}) dz = \sum_n \frac{E_n(\vec{r}_\perp)}{k_n^2} \left[ \frac{\partial}{\partial s} + \frac{\partial}{c \partial t} \right] g_n(z_0 - s, t_0) =$$

$$= \sum_n \frac{E_n(\vec{r}_\perp)}{k_n^2} \left[ \frac{g_n(z_0 - s + 0.5\Delta s, t_0 + 0.5\Delta t) - g_n(z_0 - s - 0.5\Delta s, t_0 - 0.5\Delta t)}{\Delta s} \right] + O(\Delta s^2) , \quad (28)$$

where $\Delta s = c \Delta t$.

Note, that problem (27) is a two dimensional one and can be solved by an eigenvalue solver for Laplace's equation. Additionally, the sum (28) converges relatively fast due to the $O(k_n^{-2})$ behavior of the summands.

If the number of modes in equation (28) is insufficient then applying the transverse gradient operator in the Panovsky-Wenzel theorem (4) will result in a non-smooth transverse behaviour of the transverse wakepotential. A simple and efficient way to resolve this problem is to use the harmonic transverse behaviour of the longitudinal wakepotential (15). Indeed, after we have calculated the longitudinal wakepotential with the help of the waveguide mode expansion (28) in the transverse area $\Omega^\perp$ we can take the values of the longitudinal wakepotential on the boundary $\partial\Omega^\perp$ (or near to it) and solve the Laplace equation (15) in order to find the wakepotential in the interior. This will result in a smooth transverse behaviour of the longitudinal wakepotential which allows to calculate the transverse wakepotential accurately.

## V. NUMERICAL EXAMPLES

In this section we present several numerical tests which confirm the accuracy and high efficiency of the suggested indirect methods for wake potential integration.

The wakes of the LCLS round-to-rectangular transition shown in Fig. 4 are calculated by the methods of Sections IV.B, C in reference [19]. Hence, we consider here only numerical tests for the most general indirect method described in section IV.D.

As the first example we consider the round stepped collimator shown in Fig. 3 with dimensions a=4mm, b=2.5mm and c=20mm. The longitudinal wake potential for a Gaussian bunch moving along the axis with the RMS length $\sigma = 20\mu m$ is shown in Fig. 5 on the left. We compare the wakes calculated by the direct method (see, equation (2)) against the wake potential calculated by the indirect method of section III.B. The direct wakes are obtained by integration of the longitudinal electric field component $E_z$ at the radius $r = 2.5mm$ along the z-axis for different distances between 0.25 and 4 meters. This numerical check shows that the catch-up distance is more than 4 meters. The above numerical results are obtained with the code Echo [14] in a rotationally symmetric geometry. In order to check the implementation of the 3D indirect method of section IV.D we have calculated the same example with the 3D code [16]. For 3D calculations we used the same longitudinal mesh step as for the 2D code. For the waveguide mode expansion (28) we used 200 (general) modes. The comparison of 2D and 3D results is shown in Fig. 5 on the right. Additionally, we have found that the numerical results agree well with the analytical approximation for the stepped collimator [2].

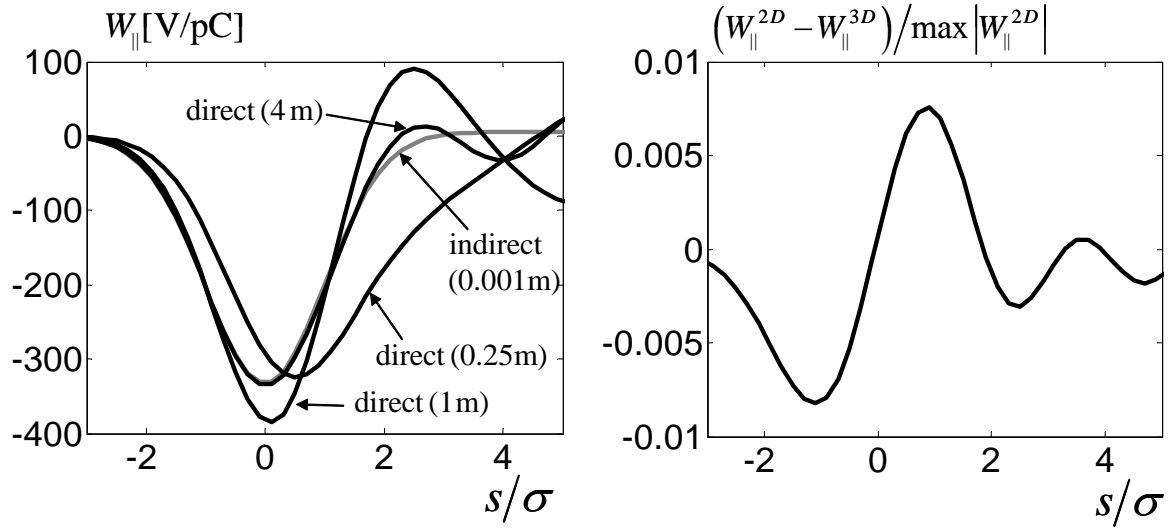

Fig. 5. Wake potentials for the round stepped collimator calculated with direct and indirect methods. Comparison of indirect wake potentials calculated with 2D (section III.B) and 3D (section IV.D) methods.

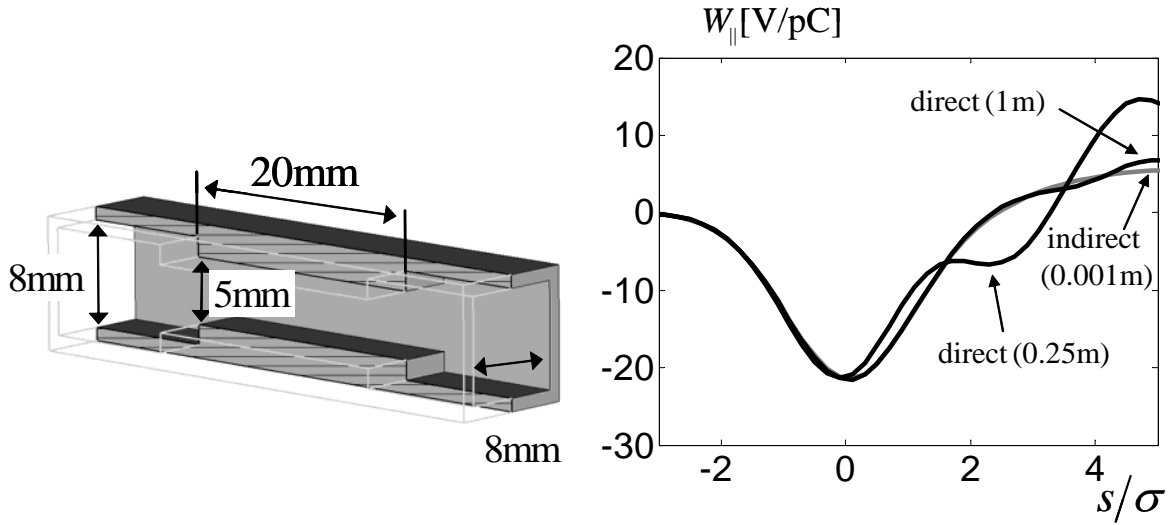

Fig. 6. Rectangular collimator and its wake potential calculated with direct and indirect (Section IV.D) methods.

As the last example we consider a rectangular collimator shown in Fig. 6. We compare the wakepotentials for a Gaussian bunch moving along the axis with the RMS length $\sigma = 200\mu m$, calculated by the direct method, and the wakepotential calculated by the indirect method of section IV.D (with 100 modes in the waveguide mode expansion). The direct wakes are obtained for distances 0.25 and 1 meter after the collimator. We again see that the indirect method applied at 1 mm after the collimator yields the accurate result, which agrees with the direct calculation at 1 meter.

## VI. CONCLUSION

In this paper we reviewed available and introduced new techniques for indirect integration of the wakepotential. The developed algorithms are checked numerically and their efficiency is confirmed by the solution of real accelerator problems [19, 21-22].

## ACKNOWLEDGEMENT

I would like to thank M. Dohlus for useful discussions and corrections. The work was supported by EUROFEL project.